\documentclass[12pt,preprint]{aastex}

\shorttitle{Dynamical structure of the 47 UMa planetary system}
\shortauthors{Ji J. et al}

\begin{document}

\title{Could the 47 UMa Planetary System be a Second Solar System: predicting the Earth-like planets}
\author{Jianghui JI\altaffilmark{1,3}, Lin LIU\altaffilmark{2,3},
Hiroshi Kinoshita\altaffilmark{4}, Guangyu LI\altaffilmark{1,3}}
\email{jijh@pmo.ac.cn}

\altaffiltext{1}{Purple  Mountain  Observatory , Chinese  Academy
of  Sciences,  Nanjing  210008, China}
\altaffiltext{2}{Department of Astronomy,  Nanjing University,
Nanjing  210093, China}

\altaffiltext{3}{National Astronomical Observatory, Chinese
Academy of Sciences,Beijing 100012,China}
\altaffiltext{4}{National Astronomical Observatory,
Mitaka, Tokyo 181-8588,Japan}

\begin{abstract}
We numerically investigated the dynamical architecture of 47 UMa
with the planetary configuration of the best-fit orbital solutions
by Fischer et al. We systematically studied the existence of
Earth-like planets in the region 0.05 AU $\leq a \leq 2.0$ AU for
47 UMa with numerical simulations, and we also explored the packed
planetary geometry and Trojan planets in the system. In the
simulations, we found that "hot Earths" at 0.05 AU $\leq a < $ 0.4
AU can dynamically survive at least for 1 Myr. The Earth-like
planets can eventually remain in the system for 10 Myr in areas
involved in the mean motion resonances (MMR) (e.g., 3:2 MMR ) with
the inner companion. Moreover, we showed that the 2:1 and 3:1
resonances are on the fringe of stability, while the 5:2 MMR is
unstable. Additionally, the 2:1 MMR marks out a remarkable
boundary between chaotic and regular motions, inside, most of the
orbits can survive, outside, they are mostly lost in the orbital
evolution. In a dynamical sense, the most likely candidate for
habitable environment is  Earth-like planets with orbits in the
ranges 0.8 AU $\leq a < 1.0$ AU and 1.0 AU $ < a < 1.30$ AU
(except 5:2 MMR and several unstable cases) with relatively low
eccentricities. The Trojan planets with low eccentricities and
inclinations can secularly last at the triangular equilibrium
points of the two massive planets. Hence, the 47 UMa planetary
system may be a close analog to our solar system, bearing a
similar dynamical structure.
\end{abstract}


\section{Introduction}
The main sequence star 47 UMa is of spectral type G0 V with a mass
of 1.03$M_{\odot}$. Butler \& Marcy (1996) reported the discovery
of the first planet in the 47 UMa system which has become one of
the most eye-catching systems particularly after the subsequent
release of an additional companion (Fischer et al. 2002). It is
sometimes thought to be a close analog of our own solar system:
for example, the mass ratio of the two giant companions in 47 UMa
is $\sim$ 2.62 (see Table 1), as compared to that of
Jupiter-Saturn (JS) of 3.34; and the ratios of two orbital periods
are close to each other. Following the analogy,  one may wonder
whether there exists additional members in 47 UMa system.
Multi-planet systems seem to be common in nature (Fischer et al.
2003; hereafter Paper I) more than ten of such systems have been
detected in Doppler surveys to date. Moreover, if the solar system
is a typical one, how are we to understand the analogous
structures in other multi-planet systems, e.g., the presence of
the inner low-mass terrestrial planets, the asteroidal belts or
Kuiper belts (e.g. \textit{Spitzer} Mission) around central stars?
The pioneering numerical experiments (Adams \& Laughlin 2003)
showed that planet-planet scattering can yield  planetary orbits
with $a\sim 1$ AU for a packed system consisting of 10 planets
with widely spaced orbits. However, from an observational
viewpoint, because of the variability of stellar atmospheres, the
current ground-based observational precision $\sim$ 2-3 m s$^{-1}$
(Butler et al. 2003) indeed places a fundamental limit on the
discovery of Earth-mass planets\footnote{however, the recent
discovered Neptune-mass planets with close-in orbits ($a<0.1$ AU)
(Butler et al. 2004; McArthur et al. 2004; Santos et al. 2004) are
grouped as a new category linking giant gaseous planet and
Earth-mass planet.} around 1 AU (even smaller), and the preference
detection of planets with masses of several $M_{\oplus}$ should
benefit from future space projects that can carry out high
resolution astrometric measurements (e.g., SIM and GAIA).
Nevertheless, in the current stage, this leaves room for
researchers to study the dynamical structure of such planets or
their formation  that may possibly reside in the systems, in
advance of the observing missions.

\section{Formation scenario}
The planetary formation scenarios may be replayed as follows: in
the standard model of core-accretion scenario (Safronov 1969;
Lissauer 1993), solid cores (or planetary embryos) rapidly grow
larger by the accretion of kilometer-sized planetesimals via
runaway growth (Kokubo \& Ida 1998; Ida \& Lin 2004), leading to
the formation of the gaseous giant planets through the accretion
of disk gas onto Earth-mass solid cores and more planetesimals
into an envelope. When they formed outside the ice boundary $\sim
3$ AU (Ida \& Lin 2004) in the system, two giant planets may
migrate inward through disk-planet interactions (Ward 1997; Kley
2000; Nelson et al. 2000; Papaloizou 2003), and such migration
process may not halt until the two planets are eventually captured
into $\sim$ 8:3 mean motion resonance (Laughlin, Chambers \&
Fischer 2002; hereafter Paper II), see also the 2:1 resonant
capture scenario for two massive planets in GJ 876 (Lee \& Peale
2002; Kley, Peitz, \& Bryden 2004). In the migration scenario, the
giant protoplanets may clear up wide gaps about their orbits
(Goldreich, Lithwick, \& Sari 2004) and cease accreting small
bodies. But the giant planets may make the residual unaccreted
planetesimals in the debris disk move inward or outward through
gravitational scattering mechanism, even with the eccentricities
and inclinations of these planetesimals  excited by sweeping
secular resonance (Nagasawa \& Ida 2000), thus the orbits of the
small bodies can undergo mutual crossings and then they are either
directly cleared up or collided into fragments in the
post-formation stage. Another possible fate for these isolated
members is that they may be captured into mean motion resonances
with the giant planets under migration (e.g., the capture of the
KBOs in the Neptune migration [Malhotra, Duncan, \& Levison
2000]), and such resonant swarm may outline the structure of
asteroidal belts in the system. In consequence, a dynamically
quasi-stable configuration with spaced orbits may be achieved to
support further secular evolution, which may determine the
resulting destinies of those members, leading to the final
assemblage of the planetary system (Lin 2004).

Paper II and Gozdziewski (2002) studied the long-term stability of
47 UMa and pointed out that the secular apsidal resonance can help
stabilize the two giant planets in an aligned configuration with
the libration of their relative periapse longitudes (Ji et al.
2003), then the eccentricities are well maintained to free from
larger vibrations due to this mechanism, as a result, this system
can even survive for billion years (Barnes \& Quinn 2004). Several
pioneer works were concentrated on the structure of the system and
presented a preliminary understanding of this issue. Jones, Sleep
\& Chambers (2001) investigated the existence of Earth-mass
planets in the presence of one known giant planet, and
subsequently Paper II  and Asghari et al. (2004) further studied
the stability of massless test particles about the so-called
Habitable Zones (HZ) according to some earlier solutions (Fischer
et al. 2002), where the dynamical model was treated as a
restricted multi-body problem. Nevertheless, as the terrestrial
planets possess significant masses, they can interact with the two
giant planets by mutual gravitation, which may result in secular
effects for the planetary system. Accordingly, we should take into
account the masses of terrestrial bodies in the model when
exploring the dynamical architecture. In this paper, we performed
extensive simulations to examine the dynamical architecture in
both the HZ and extended areas, for Earth-like planets (with
masses from 0.1 $M_{\oplus}$ to 10 $M_{\oplus}$) of 47 UMa with
stable coplanar planetary configuration, based on the best-fit
orbital parameters given by Paper I. These new reliable orbital
solutions are derived from additional follow-up observations,
hence they can represent the actual motions of the system under
study. On the other hand, as mentioned previously, the discovery
of three close-in Neptune-mass planets demonstrates that it may be
possible for less massive planets ($\sim M_{\oplus}$) to move
close to the star. Therefore, in the extended study, we also
explored low-mass planets in the region 0.05 AU $\leq a < 0.4$ AU
and we found that the secular resonance arising from the inner
giant planet can render the eccentricity excitations for the
Earth-like planets (see \S3.2). In addition, we also carried out
two other runs to predict potential bodies in 47 UMa : In \S3.1,
to compare with the inner solar system, we further investigated
the planetary configuration of 2 giant planets plus 4 terrestrial
planets, and in \S3.3 we examined the case of presence of Trojan
planets with respect to the two giant planets. The results suggest
that the 47 UMa system may have a similar architecture to the
solar system.

In the simulations, we use an N-body codes (Ji, Li \& Liu 2002) of
direct numerical simulations with the RKF7(8) and symplectic
integrators (Wisdom \& Holman 1991). We always take the stellar
mass and the minimum planetary masses from Table 1. The adopted
time stepsize is usually $\sim$ 1\%-2.5\% of the orbital period of
the innermost planet, which is sufficiently small for the
integration. Additionally, the numerical errors were effectively
controlled over the integration timescale, and the total energy is
generally conserved to $10^{-6}$ - $10^{-8}$  for the
integrations. Our main results now follow.

\section{Simulations Results}
\subsection{2 Giants plus 4 terrestrial planets}
The  announcement of the fourth planet in the 55 Cancri system
(McArthur et al. 2004) suggests that multiple planetary systems
resembling our solar system could be quite common in the galaxy.
In the first runs, we examine the configuration consisting of 2
giant planets (2G) and 4 terrestrial planets (4T) to study the
coexistence of multiple bodies. Specifically, this means that we
directly place Mercury, Venus, Earth and Mars into the 47 UMa
system to simulate "the inner solar system", where the orbital
elements for above terrestrial planets are calculated from JPL
planetary ephemerides DE405 at Epoch JD 2448750.9 corresponding to
the outer companion (see Table 2), e.g., the semi-major axes are
respectively, 0.387, 0.723, 1.00 and 1.523 AU. The giant planets
are always assumed to be coplanar in the simulations, thus the
inclinations for 4T refer to the fundamental plane of the 2G's
orbits. In Figure 1a, the orbital evolution of 4T is shown, where
$Q=a(1+e)$, $q=a(1-e)$ are, respectively, the apoapsis and
periapsis distances. The simulation results indicate that the
orbits of these low-mass planets become chaotic after $2\times
10^{5}$ yr: Mars leaves its initial orbit in less than $10^{4}$ yr
and its $Q$ can amount to $\sim 10^{3}$ AU at $t>1.5\times 10^{5}$
yr. Such dynamical instability stems from strong perturbation by
47 UMa b (see also \S3.2.4), and the Hill radius is
$R_{H}={\left[{M_{1}}/(3M_{c})\right]}^{1/3}a_{1}$ ($M_{c}$, the
stellar mass; $M_{1}, a_{1}$, the planetary mass and semi-major
axis of Companion B), then $3R_{H}\simeq 0.6$ AU, showing that the
orbit of Mars is quite close to the outskirts of $3R_{H}$ sphere.
In addition, as of  its eccentricity grows, Venus begins to cross
the orbits of Earth and Mercury at $t>2.0\times 10^{4}$ yr, which
leads to eventual destruction of the system.

However, as the two giant planets in 47 UMa are much closer to the
central star than the JS-pair to the Sun, for perfect analogy with
the inner solar system, the semi-major axes of 4T should be
shifted to 0.15, 0.29, 0.40 and 0.61 AU. Bearing this in mind, we
restarted a new run for the 2G-4T system, where we adopted the
rescaled semi-major axes for 4T together with all the other
initial values in Table 2. In this numerical experiment, we found
that the 2G-4T system can be dynamically stable and can last at
least for 20 Myr (see Figure 1b), where the time behaviors of $Q$
and $q$ for 4T show regular motions that their semi-major axes and
eccentricities do not dramatically change in their secular orbital
evolution. More terrestrial planets can be possibly created in the
later stage of planetary formation, Chambers (2001) studied the
accretion of $\sim150$ planetary embryos with lunar-to-Mars masses
in giant planets systems and recovered 4T similar to those in the
solar system. However, Levison \& Agnor (2003) underlined that the
population and masses of the resulting terrestrial planets may be
affected by the giant planets, because the growth and evolution of
the planetary embryos are determined by the possibly experienced
dynamical mechanisms (e.g., secular resonances).

\subsection{Terrestrial planets in Habitable zones}
The Habitable Zones are generally conceived as places where the
biological evolution of life is able to develop on planetary
surfaces in environment of liquid-water, subtle temperature and
atmosphere components of CO$_{2}$, H$_{2}$O and N$_{2}$ (Kasting
et al. 1993); at the same time, the planetary habitability is also
related to the stellar luminosity and the age of the star-planet
system (Cuntz et al. 2003). The HZ could be considered to be
centered at $\sim$ 1 AU $(M_{c}/M_{\odot})^{2}$. For 47 UMa, the
inner and outer boundaries of HZ range from 0.7 AU to 1.3 AU
(Menou \& Tabachnik 2003), however, in our practical simulations,
we extended the HZ to other areas for the purpose of a more
comprehensive study.

In the second series of runs, we extensively investigated the case
of two giant companions with one terrestrial planet in the HZ. The
mass of the assumed terrestrial planet ranges from 0.1
$M_{\oplus}$ to 10 $M_{\oplus}$. And the adopted initial orbital
parameters are as follows: numerical scanning was carried out for
[$a, e$] space by direct integrations, where the low-mass bodies
were placed at equal intervals of 0.01 AU for 0.05 AU $\leq a \leq
2.0$ AU, the eccentricities were uniformly spaced every 0.01 for
$0.0 \leq e \leq 0.2$ (for 0.05 AU $\leq a < 0.4$ AU, where $0.0
\leq e \leq 0.1$), the inclinations are $0^{\circ} < I <
5^{\circ}$, and the other angles were randomly distributed between
$0^{\circ}$ and $360^{\circ}$. Thus, over 3000 simulations were
exhaustively performed for  typical integration spans from 1 Myr
to 10 Myr, totalling several $10^{10}$ yr.

\subsubsection{0.05 AU $\leq a < 0.4$ AU}
In these runs, we explored the secular evolution of 385 "hot
Earths" or "hot Neptunes" for a time span of 1 Myr. All the
simulations are dynamically stable for $10^{6}$ yr, and 96\% of
the orbits bear  $e_{final}<0.20$. However, Figure 2a shows that
the eccentricities for the bodies at $\sim 0.30$ AU are excited to
$\sim 0.40$, where the secular resonance $\nu_{1}$
($41^{"}.11$/yr) of the inner companion (similar to $\nu_{5}$ for
Jupiter) is responsible for the excitation of eccentricity. The
debris disk at $\sim 0.30$ AU is also shown by Malhotra (2004),
who presented similar results of the eccentricity excitation of
massless bodies by nonlinear analytic theory for secular
resonance. In addition, even more terrestrial planets with spaced
orbits can simultaneously survive for longer lifetime (e.g., the
second simulation of 2G-4T system in \S3.1). Nevertheless, one can
easily see that this region is not a good location for
habitability, due to extra high temperature. Recently, Narayan,
Cumming, \& Lin (2005) discussed the detectability of low-mass
objects with orbital periods $P \leq 10$ days, and they estimated
the present threshold for detecting planets with close-in orbits
is $\geq 10-20 M_{\oplus}$, depending on the number of the
observations and the precision of Doppler measurements. Indeed,
the improvement of precision of the ground-based observations will
lead to the discovery of additional low-mass planets ($\sim 10
M_{\oplus}$) in other known planetary systems (G. W. Marcy 2004,
private communication).

However, it brings about great difficulty in catching the
planetary formation for hot planets that move so close to their
host stars. Although the gas giant planets may undergo inward
orbital migration from several AU to the vicinity of host stars,
Ida \& Lin (2004) found that a large fraction (90\%-95\%) of the
planets that have migrated to $a <0.05$ AU must perish, because
the tidal heating can make the planetary radius inflate (Gu, Lin,
\& Bodenheimer 2003) and the body may be directly engulfed by the
star (Israelian et al. 2001) or survive only as rocky "cores" with
the gas envelope removed due to the mass loss of the planet
through the material exchange at the Roche radius. Nevertheless,
it is hopeful to detect such planets with $a \ge 0.05$ AU in
future space missions (e.g., COROT, KEPLER, TPF), while the
survival of the short-period terrestrial planets (Mardling \& Lin
2004) is also related to the relativistic potential of the star.

\subsubsection{0.4 AU $\leq a < 1.0$ AU}
We carried out 1260 integrations in this region for 5 Myr and we
found that none of the orbits escaped during this time span and
94\% of them were in the resulting $e < 0.25$, see the final
status shown in Figure 2b. We find that the eccentricities of the
orbits with 0.70 AU $< a < $ 0.78 AU can be pumped up and in the
2:9 MMR at $\sim 0.76$ AU, $e$ can reach $\sim 0.90$, indicating
that there may exist a gap near this resonance. Most of the
Earth-like planets about 1:4 MMR at $\sim 0.82$ AU move stably in
bounded motions with low-eccentricity trajectories, except for two
cases where the eccentricities eventually grow to high values.
Paper II pointed out that the secular resonance $\nu_{2}$ arising
from the outer companion (similar to $\nu_{6}$ for Saturn) can
remove the test bodies. Would the $\nu_{2}$ also influence the
Earth-like planets in this system? Nevertheless, we did not find
this mechanism at work at about $\sim 0.85$ AU (see Paper II) when
we examined the results, because the terrestrial planets under
study that all bear finite masses that may change the strength of
this resonance; on the other hand, the location of the secular
resonance is changed due to the orbital variation of the outer
companion. For a terrestrial planet with a mass of $10
M_{\oplus}$, the region for $\nu_{2}$ secular resonance is now
shifted to $\sim 0.70$ AU, where two eigenfrequencies for the
terrestrial body and outer giant planet given by the
Laplace-Lagrange secular theory are, respectively, $211^{"}.37$/yr
and $225^{"}.48$/yr. This indicates that both planets almost have
the same secular apsidal precession rates in their motion. At the
new location, the $\nu_{2}$ resonance, together with the mean
motion resonance, can work at clearing up the planetesimals in the
disk (see Fig. 2b) by the excitation of the eccentricity;
qualitatively, our results are in accord with those of Paper II.

While the inner edge of HZ marks out a narrow unstable area, it is
very possible to discover Earth-like planets in this wider area,
0.4 AU $< a < 1.0$ AU, in future surveys, which are to be the best
candidate of habitable places for biological evolution of
intelligent beings.

\subsubsection{1.0 AU $\leq a < 1.3$ AU}
There were 630 simulations in this region for 10 Myr and we found
that 88\% survived the integration, confirming the results given
in Paper II that most of the test particles with $a < 1.3$ AU can
eventually remain in the system. Here, for the 3:1 resonance at
$\sim 1.0$ AU, Figure 3a shows that there are stable orbits in the
zones about 1 AU with $e\leq0.1$, which agrees with the work by
Rivera \& Haghighipour (2004) who showed that a test particle can
last 100 Myr at 1 AU in 47 UMa; while for $0.1<e\leq 0.2$, the
orbits tend to be in unstable state owing to the excitation of the
eccentricities. The simulations may imply that the Earth-mass
planets near 3:1 resonance are possibly on the edge of stability.
However, the previous studies on this system showed that the 3:1
resonance is a gap with no survivors. Let us mention that such
differences may arise from the adopted initial planetary
configurations, and here we adopt the reliable best-fit orbital
solutions given in Paper I that can describe the exact motions for
the two giant planets. Hence, a comparative run was carried out to
examine this, again we ran 630 simulations for 10 Myr but with the
earlier orbital elements for the two massive planets  to reproduce
the previous results at the 3:1 resonance. Our results with the
earlier data show that most of the Earth-mass planets about 3:1
resonance are unstable for the investigated time, and their
eccentricities can be pumped up to $\sim 1$ through resonance;
besides, the inclinations are excited to high values ranging
$90^{\circ}$ to $180^{\circ}$, indicating that the orbits of the
Earth-mass planets become retrograde in the dynamical evolution
and cross those of the prograde giant planets before they
terminate their dynamical lifetimes. Thus, we may safely conclude
that the stability of the terrestrial planets is dependent on the
initial planetary configuration.

In Fig.3a, a narrow unstable stripe appears at the 5:2 MMR at
$\sim 1.13$ AU, although several of them can be luckily left
behind, most of the Earth-size planets are removed at $\sim$ 1 Myr
due to the perturbation of 47 UMa b, and in this sense it is
analogous to the situation in the solar system. However, a wider
area between 3:1 and 5:2 MMR is assumed to be a qualified
candidate habitable environment where the Earth-mass planet will
not encounter the problem of dynamical stability, and this is
almost true for the region (1.13 AU, 1.30 AU) with $e\leq0.1$,
except several unstable islands near 7:3 MMR at 1.18 AU. The
smaller eccentricity (near-circular orbits) may not cause dramatic
variations of temperature on the planet's surface, so favoring
habitability. Therefore, in a dynamical sense, if the 47 UMa
system can be adopted as a candidate target for SIM, it is also
possible to detect other Earths with stable orbits about 1 AU.

\subsubsection{1.3 AU $\leq a \leq 1.6$ AU}
We performed 651 simulations for 10 Myr, and found the dynamical
structure in this regime to be quite complicated: 14\% of them can
finally survive for this time span, and 86\% are lost by ejection
into hyperbolic trajectories, indicating the chaotic nature for
these bodies. In Figure 3b, we can notice that the 2:1 MMR region
is at $\sim 1.31$ AU and also close to the outer edge of HZ, and
the orbits with $0.0 < e< 0.10$ are unstable, while for $0.10 \leq
e\leq 0.20$, there are stable islands where fictitious planets can
remain in bounded motions in the final system. We observe that the
2:1 resonance marks out a remarkable boundary between chaotic and
regular orbits, indicating that orbits with $a < 1.31$ AU can have
much larger surviving rates than those of $a > 1.31$ AU. However,
there are wider stable region about 9:5 MMR at $\sim 1.40$ AU for
low eccentricities $0.0<e\leq0.05$. Most of the unstable orbits
are in the region $1.43$ AU $<a <1.56$ AU, using resonance
overlapping criterion (Murray \& Dermott 1999), the separation in
semi-major axis
$\Delta{a}\approx1.3{\left({M_{1}}/M_{c}\right)}^{2/7}a_{1}\doteq0.496$
AU, thus, the inner boundary $R_{O}$ for 47 UMa b is at $\sim
1.58$ AU, and the orbits in this zone become chaotic during the
orbital evolution because the planets are both within $3R_{H}$ and
also close to $R_{O}$. And the characterized ejecting time $\tau
\sim 1$ Myr, which means the apparent gap (e.g., 5:3 MMR at $\sim
1.48$ AU) in the inner belt, except for several stable islands.
Another possible population for terrestrial planets is located at
3:2 resonance $\sim 1.59$ AU for $0.04< e < 0.20$, and 18 small
bodies can last for 10 Myr and confirm the results of Paper II.
The 3:2 MMR zone in 47 UMa is reminiscent of  the Hilda asteroids
in the solar system moving in a stable region.

\subsubsection{1.6 AU $< a \leq 2.0$ AU}
840 Earth-like planets are placed in this region for integration
for 10 Myr, and the simulation revealed that 98\% are removed from
the system within a typical ejection time $\tau<3\times 10^{4}$
yr, which is much shorter than for 1.3 AU $\leq a \leq $1.6 AU,
implying a thoroughly chaotic situation in this area. It is not
difficult to understand that these terrestrial planets are
entirely thrown out by the gravitational influence of the inner
giant planet, using the $3R_{H}$ stability criterion.

\subsection{Trojan planets}
In the solar system, at present there are 1690 Jupiter
Trojans\footnote{http://cfa-www.harvard.edu/iau/lists/JupiterTrojans.html}
(1062 preceding and 628 trailing Jupiter) located at L4 and L5.
Hence, a fascinating issue is that whether there exist "extrasolar
Trojan planets" in other planetary systems, and if such planets
can occur in the system harboring only one gas giant with $a \sim
1 $ AU (Dvorak et al. 2004), it may essentially help understand
their presence in the HZ. Laughlin \& Chambers (2002) found the
equilateral configuration consisting of a star and two equal-mass
planets can be linearly stable for mass ratios less than 0.03812
and showed that a pair of Saturn-mass planets can sustain the
stability in the 1:1 orbital resonance.

In the third series of runs, we let the experimental planets with
masses of 0.1 $M_{\oplus}$ to 10 $M_{\oplus}$ initially orbit
about the triangular equilibrium points of the two giant
companions, for each giant planet there are 100 orbits distributed
near L4 (or L5) point. Then, we implemented 400 dynamical
simulations to explore their secular evolution on the timescale of
$10^{7}$ yr. The starting conditions for the Trojan bodies are:
$0.00 < e < 0.10$, $0^{\circ} < I < 1^{\circ}$ and $|\lambda -
\lambda_{1,2} - 60^{\circ}|<1^{\circ}$ (where $\lambda$, and
$\lambda_{1,2}$, respectively, the mean longitudes of the Trojan,
the inner and outer companions). For the inner planet of 47 UMa b,
the survival rates for the low-mass bodies about L4 and L5 points
are, respectively, 96\% and 98\%, which is quite similar to the
Trojans and Greeks for Jupiter; while for the outer planet (47 UMa
c), the survival proportion goes down to 80\% and 70\% near L4 and
L5 region, as a comparison Saturn has no co-orbital bodies.
Obviously, the star-planet mass ratios $\mu \sim 10^{-3} < 0.0385$
(Murray \& Dermott 1999) both satisfy the linear stability
condition for the triangular points. In addition, it is also
suggested that the stable triangular region for the inner planet
is wider than that of the outer planet, and this can be easily
understood as this width depends on the value $\sim \mu^{1/2}$ of
the tadpole region.

Figure 4 displays the long-term orbital evolution for the low-mass
planets moving about L4 (\textit{black line}) and L5
(\textit{yellow line}) points for 47 UMa, where $a$ and $e$ both
perform small modulations for $10^{7}$ yr, and $\lambda -
\lambda_{1,2}$ also librate about $60^{\circ}$ and $300^{\circ}$,
respectively, with low amplitudes for the same time span.

\section{Summary and discussions}
In this work, we have systematically studied the existence of
Earth-like planets in the region for 0.05 AU $\leq a \leq 2.0$ AU
for 47 UMa by numerical simulations. In addition, we also
investigated the packed system and Trojan planets in this system.
We now summarize the main results as follows:

(1) The "hot Earths" can dynamically survive for $10^{6}$ yr for
0.05 AU $\leq a<$ 0.4 AU, and they are probably detected by
transit time variations (Agol et al. 2005; Holman \& Murray 2005)
via their interaction with the transiting planet\footnote{Laughlin
G. and his collaborators are now maintaining the website (see
http://www.transitsearch.org/) to search for transiting "hot
Jupiter" planets through the worldwide cooperative observational
campaign. Besides, the other active group from OGLE project have
detected several transiting planets
(http://sirius.astrouw.edu.pl/$\sim$ogle/) in the observations.}.
These techniques may become a vital and effective observational
strategy to discover more transiting objects around main sequence
stars.

(2) The Trojan planets with low eccentricities and inclinations
can last out at the triangular equilibrium points of the two
massive planets of 47 UMa. In this sense, it is analogous to the
case in solar system. Nevertheless, if they really exist, the
formation of these bodies is still a mystery and needs further
study.

(3) The Earth-like planets can eventually remain in the system for
10 Myr in the areas associated with mean motion resonance (e.g.,
3:2 MMR) with the inner companion. We also showed that the 2:1 and
3:1 resonances could be on the fringe of stability, but the 5:2
MMR is unstable and the bodies can be ejected as "extrasolar
comets". And this may sketch out an asteroidal belt structure
similar to the solar system. Moreover, the 2:1 MMR (near the outer
boundary of HZ at 1.30 AU) marks out a significant barrier between
chaotic and regular motions, implying that a large fraction of the
orbits inside this resonance can be survive, while most of them
are lost in the simulations outside the 2:1 resonance. Again,
considering the inner boundary of HZ and dynamical stability, we
point out that the most likely candidate for habitable environment
is  terrestrial planets with orbits in the ranges 0.8 AU $\leq a <
1.0$ AU and 1.0 AU $ < a < 1.30$ AU (except 5:2 MMR, and several
unstable cases) with low eccentricities (e.g., $0.0 \leq e \leq
0.1$). However, in our own solar system there are no terrestrial
planets from the 1:4 MMR out to Jupiter, although there are stable
orbits there.  This may suggest that although some orbits are
stable, conditions are such that terrestrial planets cannot form
so close to giant planets. Perhaps this is because runaway growth
is suppressed due to the increased eccentricities from the
perturbations of the giant planet. In 47 UMa, the corresponding
region runs from 0.82 AU on out (see Figure 2), almost completely
covering the HZ. Hence, it would be reasonable to conclude that
the only proper place to find habitable planets in this system
would be at about 0.8 AU. But this should be carefully examined by
forthcoming space measurements (e.g., SIM) capable of detecting
low-mass planets.

In a word, we can see that the 47 UMa planetary system may bear a
similar dynamical structure to our solar system, in that it can
also own several terrestrial members resembling the inner solar
system. A comparative study can be performed in other planetary
systems with two giant planets ({\' E}rdi et al. 2004; Raymond \&
Barnes 2005) to explore whether Earth-like planets can exist
there, and to locate less massive planets (Malhotra 2005, in
preparation) in general planetary systems. The dynamical
habitability is relevant to both mean motion resonance and secular
resonance for a given system. Moreover, the formation of the
terrestrial planets in the HZ is difficult in the planetary
systems that contain a giant planet with a moderate-size eccentric
orbit (Veras \& Armitage 2005), because most of the initial
material within HZ was depleted by orbital crossings; in addition,
the accretion simulations (Paper II) show that the massive
planetary embryos in the HZ should be created prior to the
formation of giant gaseous planets. Hence, more efforts should
also be directed to the understanding of the formation of
habitable terrestrial planets.


\acknowledgments{We thank the anonymous referee for useful
comments and suggestions that helped to improve the contents. We
thank Lin D.N.C., Marcy G.W. and Malhotra R. for insightful
discussions. We gratefully acknowledge the language improvement by
Kiang T. We are also grateful to Zhou Q.L. for the assistance of
workstation utilization. Most of the computations were carried out
on high performance workstations at Laboratory of Astronomical
Data Analysis and Computational Physics of Nanjing University.
This work is financially supported by the National Natural Science
Foundations of China (Grant 10203005, 10173006, 10233020) and the
Foundation of Minor Planets of Purple Mountain Observatory.}

\clearpage
\begin{deluxetable}{lll}
\tablewidth{0pt} \tablecaption{The orbital parameters of 47 UMa
planetary system (adopted from Fischer et al. [2003]). The stellar
mass is $1.03M_{\odot}$.} \tablehead{ \colhead{Parameter} &
\colhead{Companion B} & \colhead{Companion C} } \startdata
$M$sin$i$($M_{Jup}$)       & 2.86   & 1.09       \\
Orbital period $P$(days)   & 1079.2    & 2845.0    \\
$a$(AU)                    & 2.077     & 3.968     \\
Eccentricity $e$           & 0.05      & 0.00      \\
$\omega$(deg)              & 124.3     & 170.89     \\
Periastron Time (JD)       & 2452374.1 & 2448750.9
\enddata
\end{deluxetable}

\begin{deluxetable}{lllllll}
\tablewidth{0pt} \tablecaption{The orbital elements for 4
terrestrial planets at JD 2448750.9 (From DE405).}
\tablehead{\colhead{Planet} &\colhead{a(AU)} &\colhead{ecc}
&\colhead{$I$} &\colhead{$\Omega$} &\colhead{$\omega$}
&\colhead{$M$}} \startdata
Mercury &.387  &.2056  &7.00  &48.34  &29.10  &260.40 \\
Venus &.723    &.0067  &3.39  &76.70  &55.06  &253.70 \\
Earth &1.000   &.0163  &0.001 &255.71 &207.06 &123.84 \\
Mars  &1.523   &.0935  &1.850 &49.57 &286.48  &355.18
\enddata
\end{deluxetable}
\clearpage

\epsscale{1.10}
\begin{figure}
\figurenum{1} \plottwo{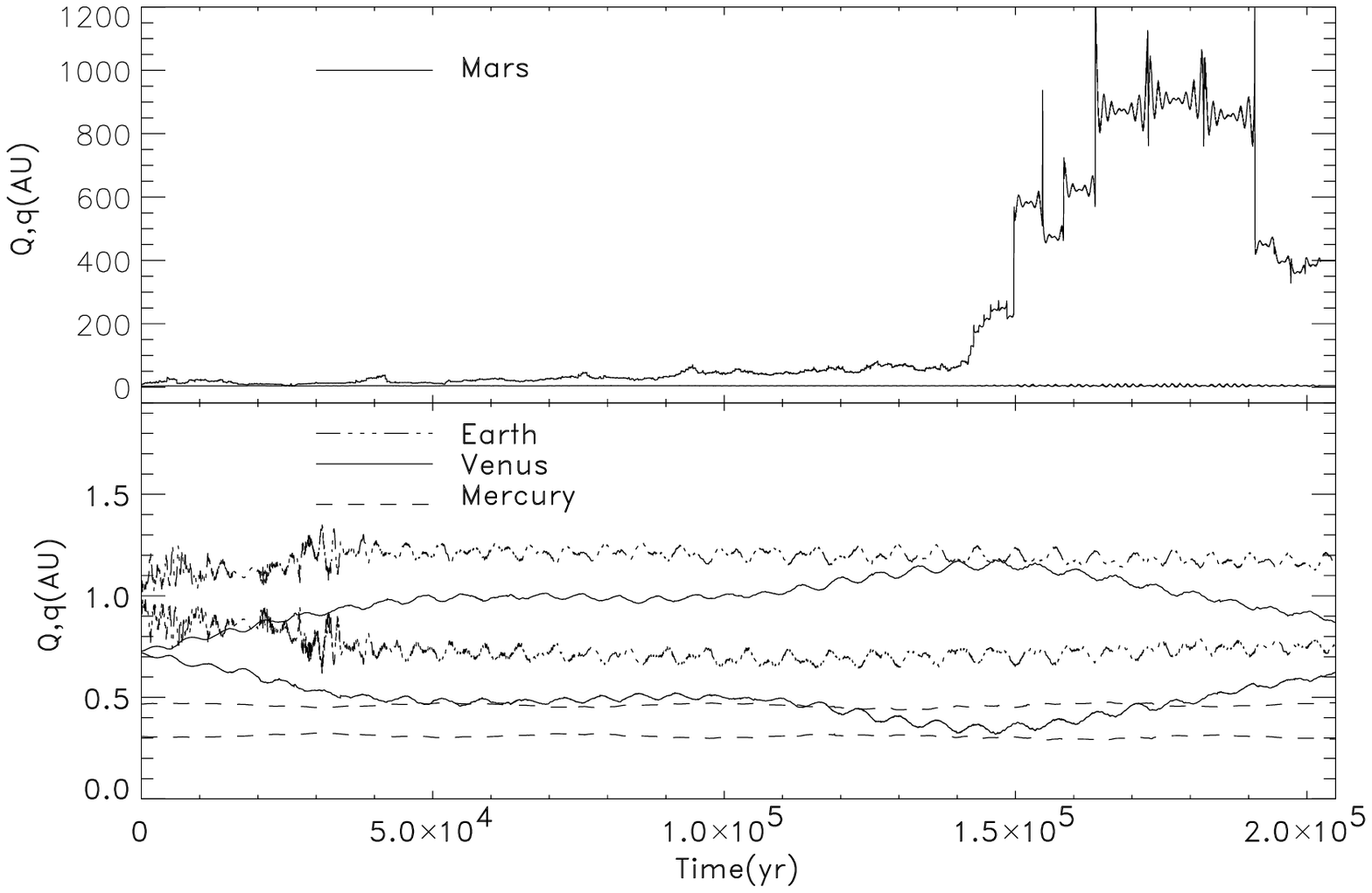}{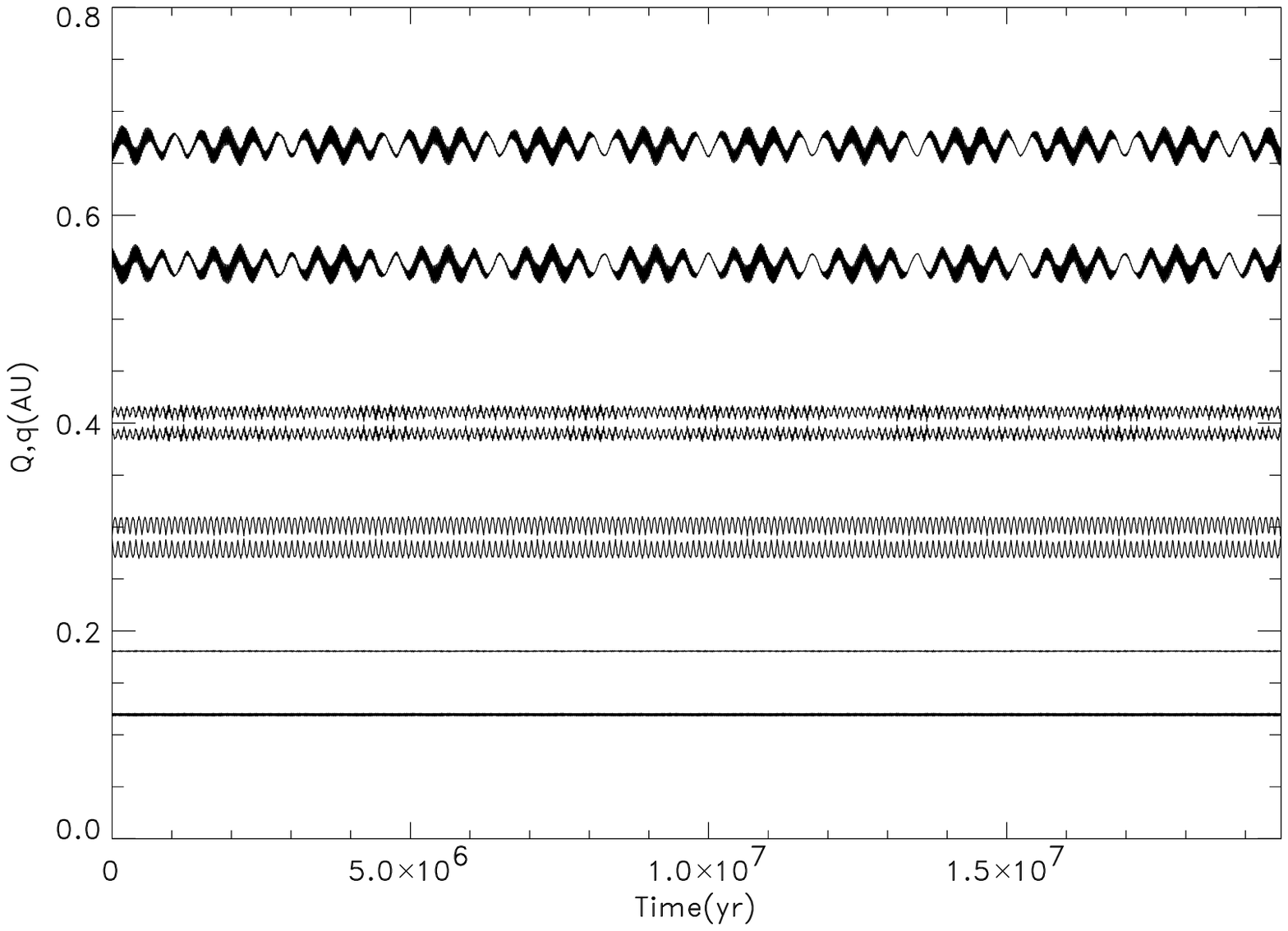} \caption{The orbital
evolution of 4 terrestrial planets. \textit{Left}: chaotic case
with $a$ from Table 2, Mars leaves its initial orbit at $t
<10^{4}$ yr, and Venus begins to cross the orbits of Earth and
Mercury at $t>2.0\times 10^{4}$ yr. \textit{Right}: with rescaled
semi-major axes, 4T remain regular motions that $a$ and $e$ both
perform slight vibrations for 20 Myr. \label{fig1} }
\end{figure}
\clearpage

\epsscale{1.10}
\begin{figure}
\figurenum{2} \plottwo{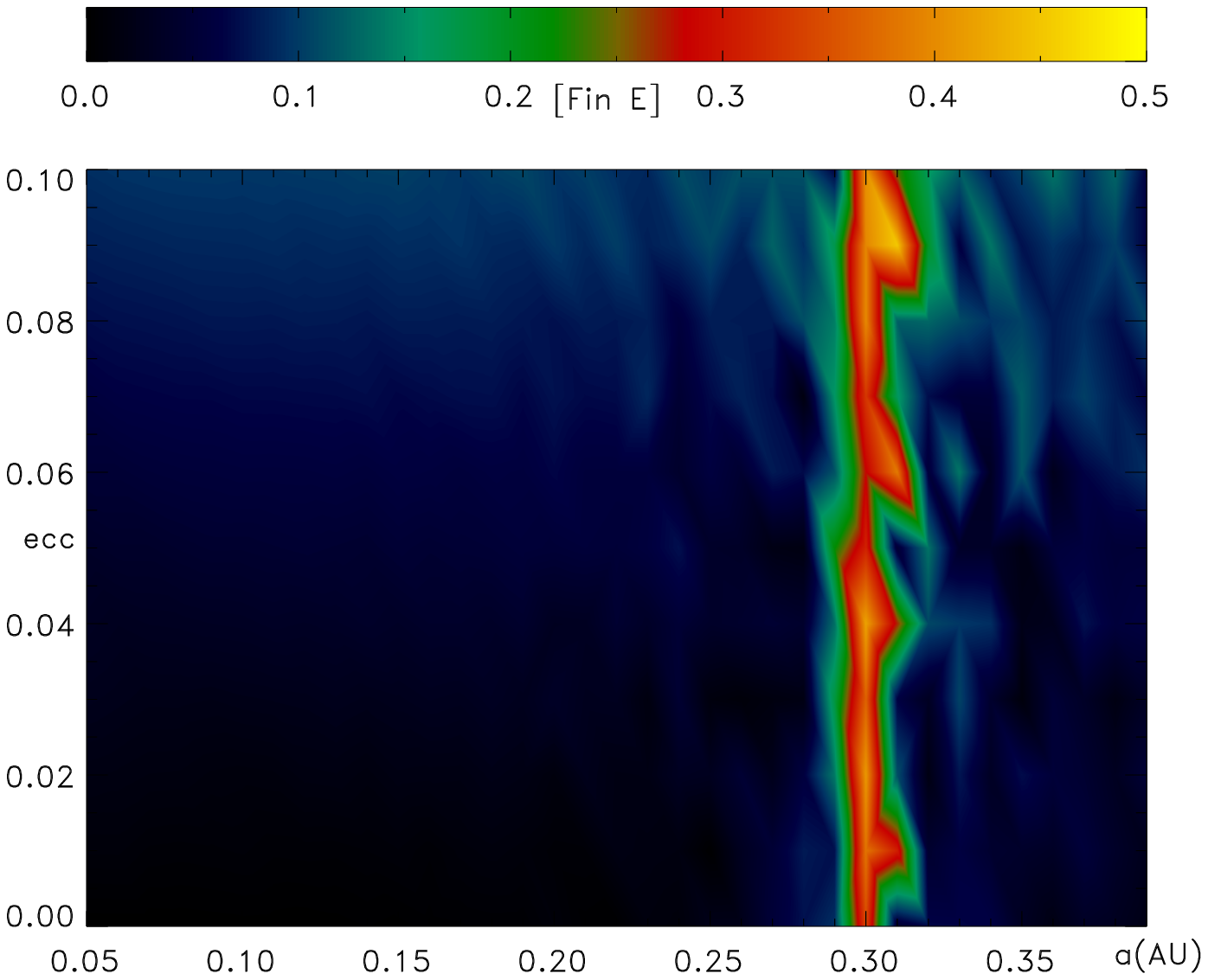}{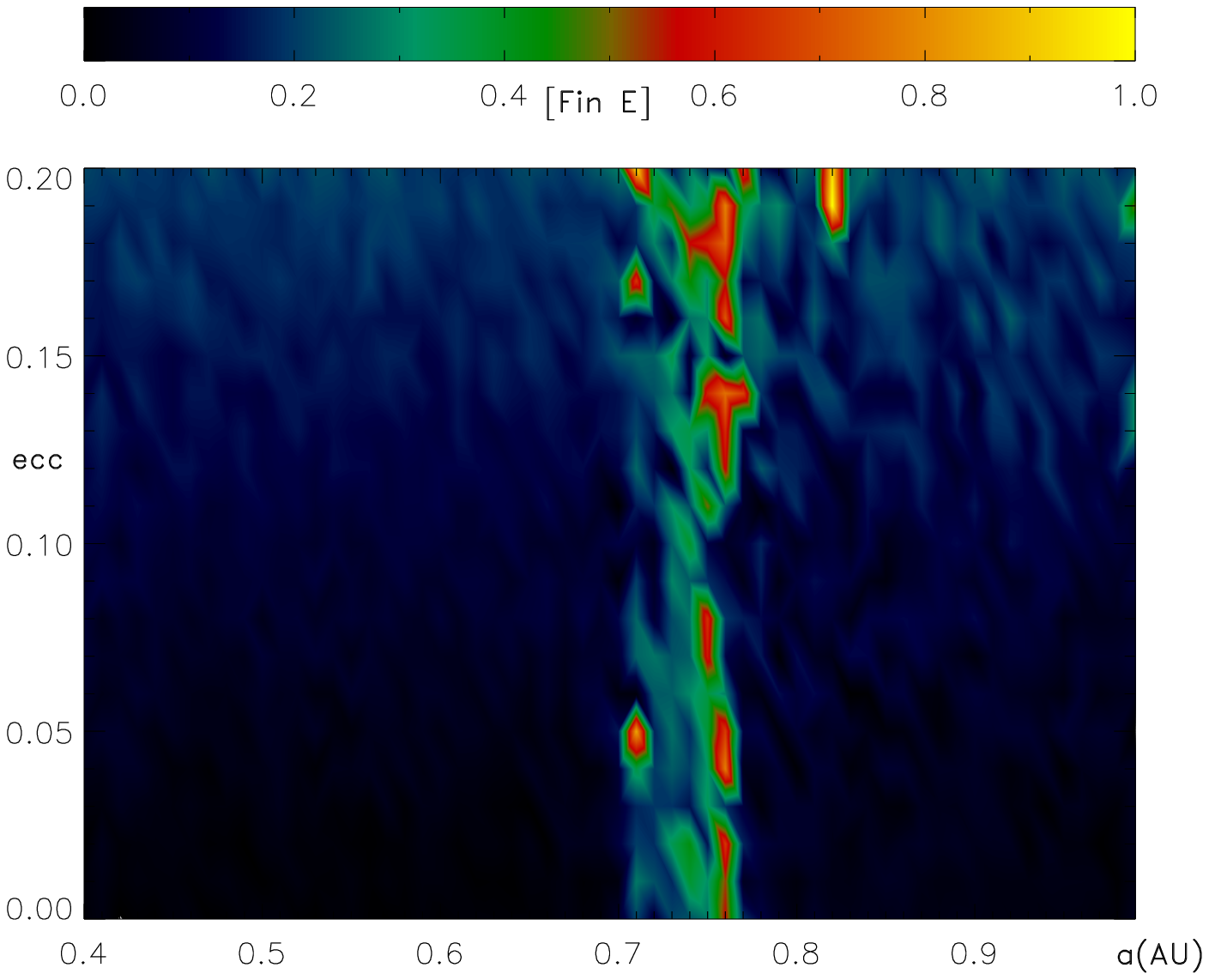} \caption{The contour of
status of the final eccentricities for Earth-like planets, the
vertical axis for the initial $e$. \textit{Left:} 0.05 AU $\leq a
< 0.4$ AU for 1 Myr.  Notice $\nu_{1}$ secular resonance at $\sim
0.30$ AU pumps up the eccentricities. \textit{Right:} 0.4 AU $\leq
a < 1.0$ AU for 5 Myr. The $e$ of the orbits with 0.70 AU $< a <
0.78$ AU can be excited and in the 2:9 MMR at $\sim 0.76$ AU, $e$
can reach $\sim 0.90$. \label{fig2}}
\end{figure}

\clearpage

\epsscale{1.10}
\begin{figure}
\figurenum{3} \plottwo{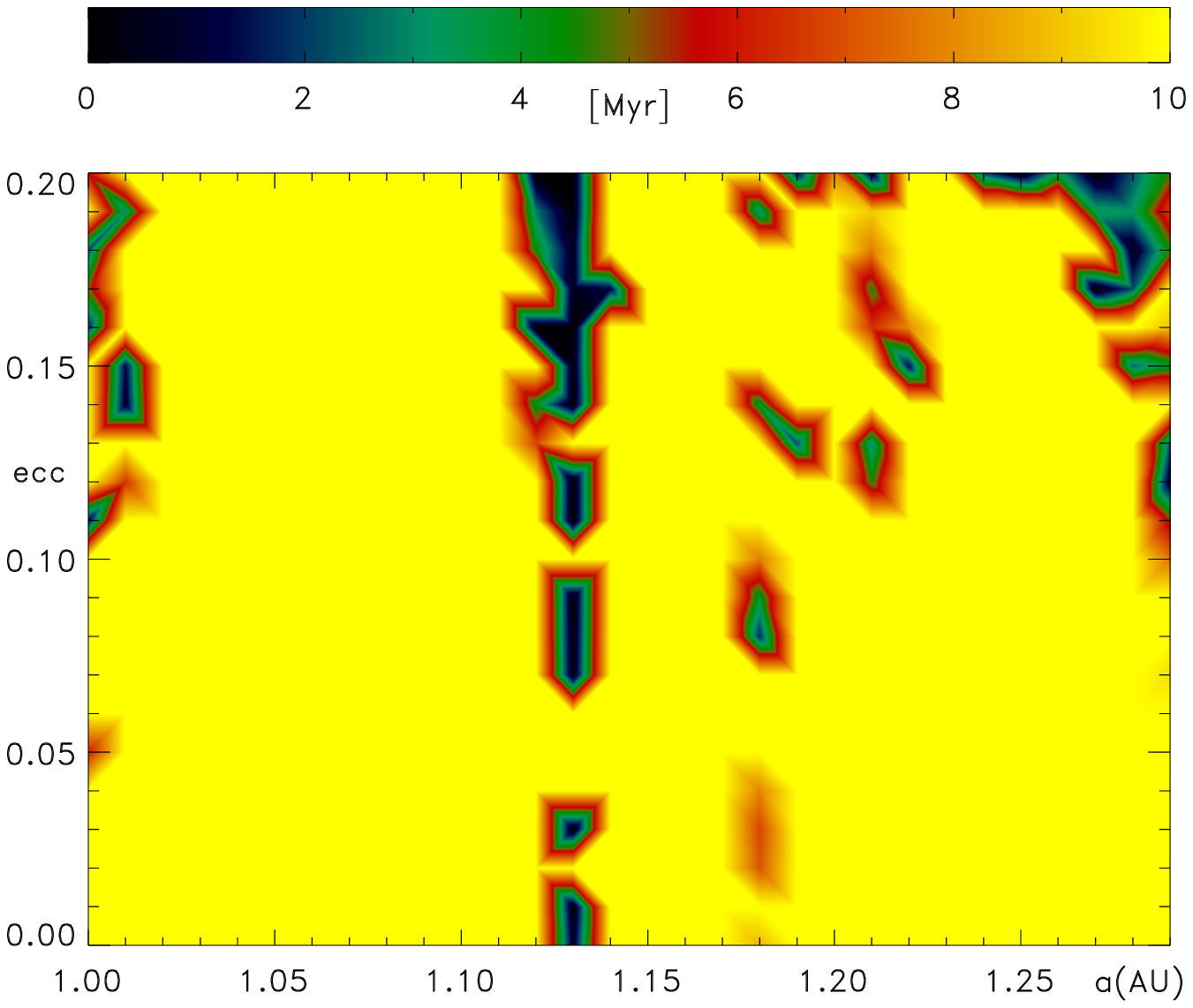}{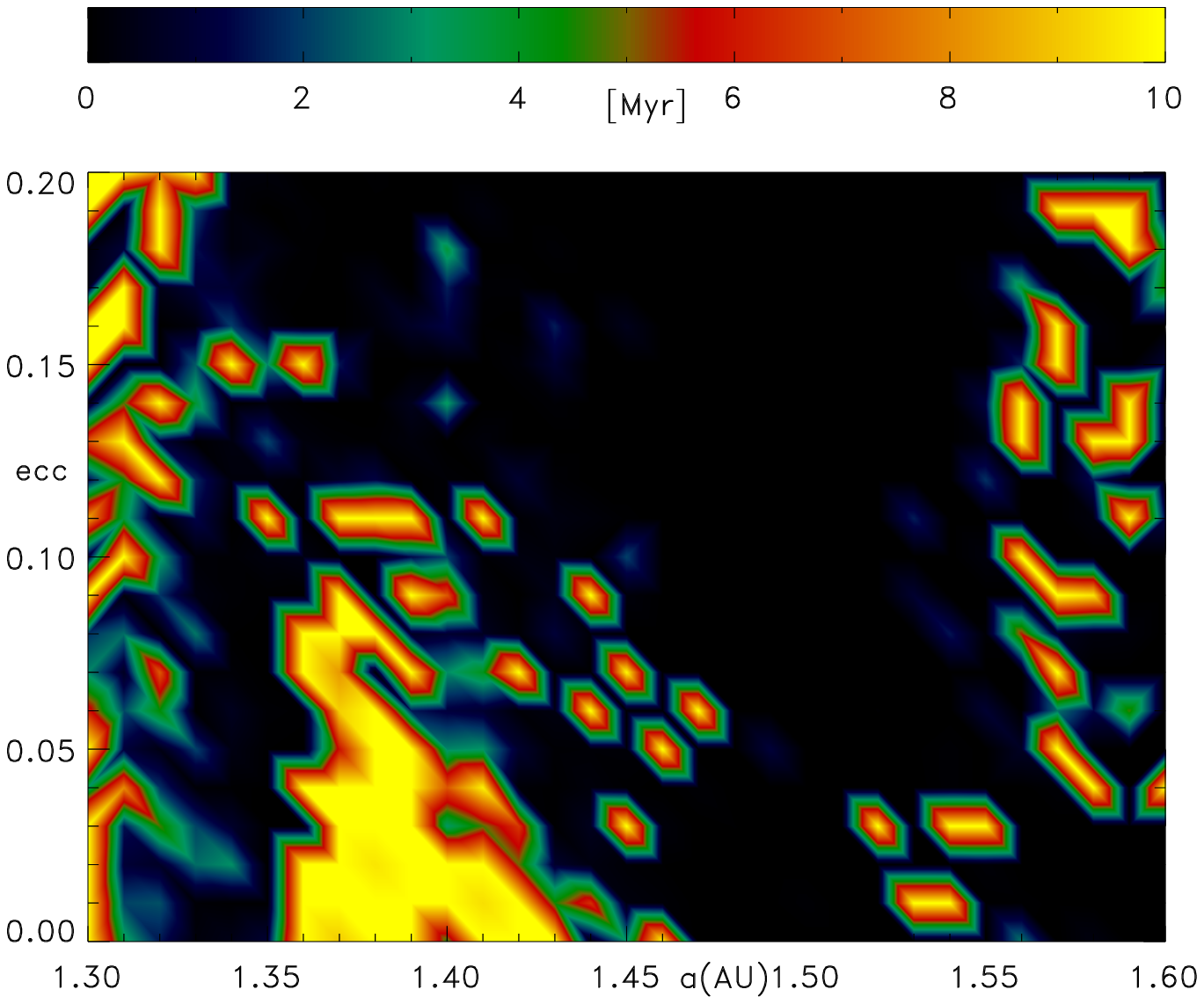} \caption{The surviving
time for Earth-like planets for the integration of 10 Myr, the
vertical axis is the same as Fig.2. \textit{Left}: for 1.0 AU
$\leq a < 1.3$ AU, see the gap for the 5:2 MMR at $\sim 1.13$ AU.
\textit{Right}: for 1.3 AU $\leq a \leq 1.6$ AU, a population of
the terrestrial planets is about 9:5 MMR at $\sim 1.40$ AU  for
low eccentricities (see texts for details). \label{fig3} }
\end{figure}

\clearpage

\epsscale{1.10}
\begin{figure}
\figurenum{4} \plottwo{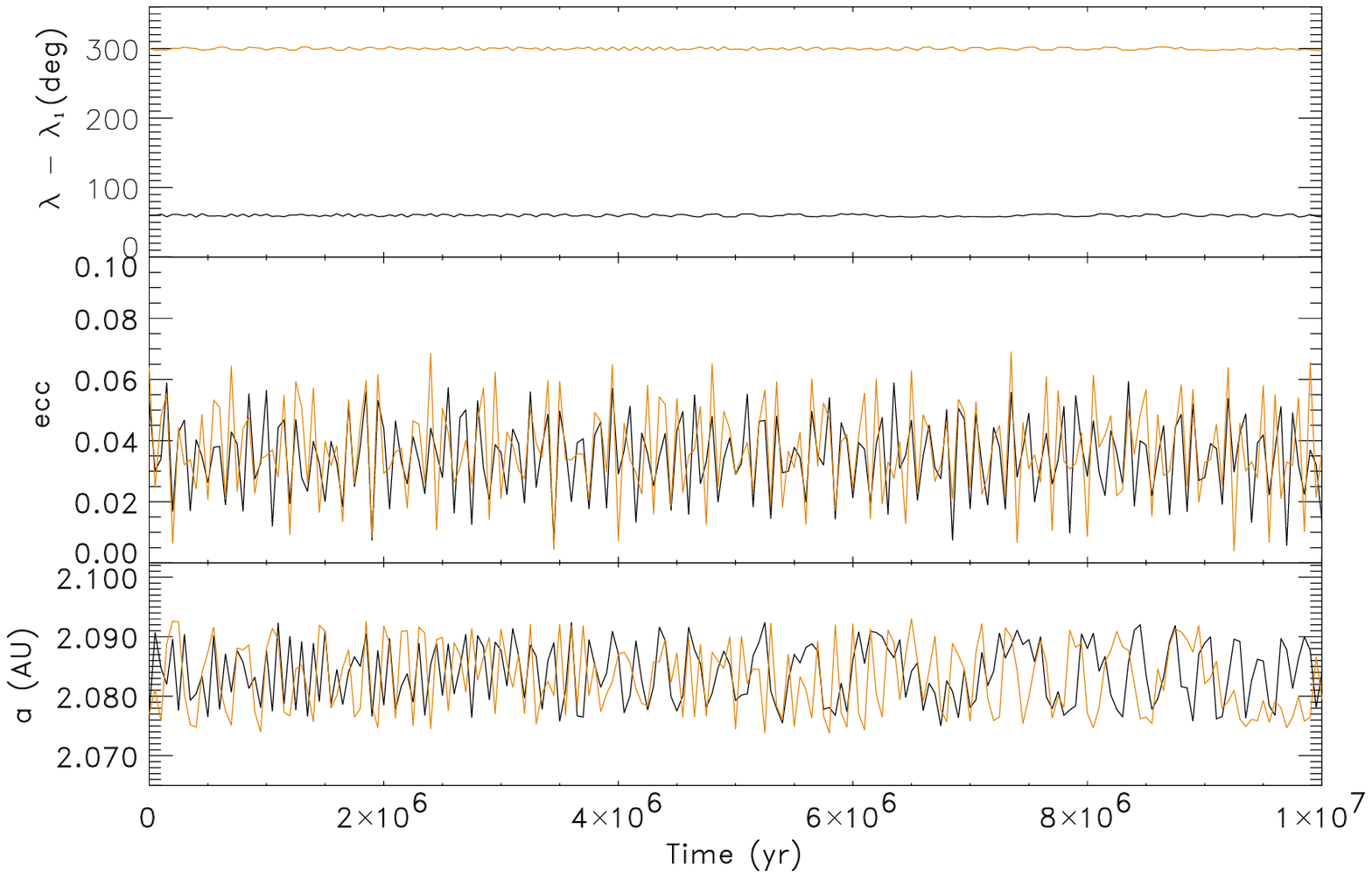}{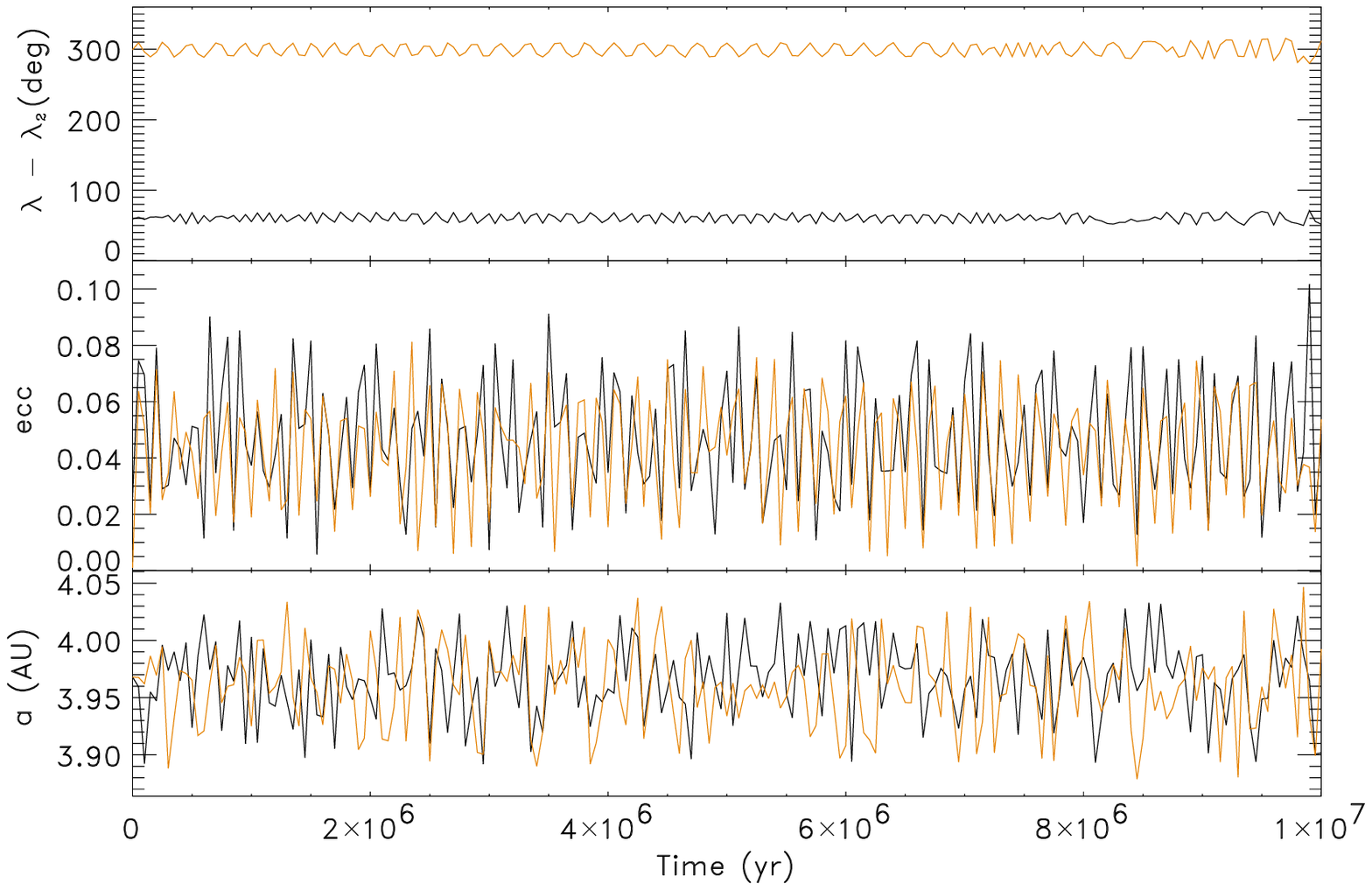} \caption{The secular
evolution for the low-mass planets moving about L4 (\textit{black
line}) and L5 (\textit{yellow line}) points for 47 UMa, where $a$
and $e$ both perform small modulations for $10^{7}$ yr, and
$\lambda - \lambda_{1,2}$ also librate about $60^{\circ}$ and
$300^{\circ}$ with low amplitudes, respectively. \textit{Left}:
for inner planet. \textit{Right}:for outer planet.\label{fig4}}
\end{figure}

\end{document}